\documentstyle[psfig]{l-aa}

\newcommand{\mdot}{\dot{M}}
\newcommand{\myr}{\ifmmode{{\rm\ M}_\odot{\rm\ yr}^{-1}}
         \else{${\rm\ M}_\odot$ yr$^{-1}$}\fi}
\newcommand{\msun}{\ifmmode{{\rm\ M}_\odot}\else{${\rm\ M}_\odot$}\fi}

\begin{document}

   \thesaurus{01         % A&A Section 1: Letters
              (02.08.1; 08.16.4; 08.05.3; 09.16.1)}

   \title{The formation of bipolar planetary nebulae}

   \subtitle{}

   \author{Garrelt Mellema
%          \inst{1}
          }

   \offprints{G. Mellema}

   \institute{Stockholm Observatory,
              S-133 36 Saltsj{\"o}baden,
              Sweden\\
              email: garrelt@astro.su.se
             }

   \date{Received ; accepted }

   \maketitle

   \begin{abstract}

Using a radiation-hydrodynamics code I follow the formation of
planetary nebulae around stars of different mass. Because a more
massive central star evolves much faster than a lower mass one, it is
to be expected that this will affect the formation of the PN.  For the
stars I use the evolutionary tracks for remnants with masses of 0.605
M$_{\odot}$ and 0.836 M$_\odot$, taken from Bl{\"o}cker (1995). The AGB
wind is assumed to be concentrated in a thin disk, which in models
without evolving stars leads to the formation of a bipolar nebula. I
find that in the case of the 0.836 M$_\odot$ remnant the nebula indeed
acquires a bipolar shape, whereas for the 0.605 M$_\odot$ remnant the
shape is more elliptical. The reason for this is the time it takes to
ionize the AGB material; if this happens sufficiently slowly the density
distribution in the AGB wind will be smoothed out, leading to more
elliptical shapes. If it happens quickly, the original density
distribution (in this case a thin disk) is hardly affected. This
result suggests that lower mass central stars will less easily produce
bipolar nebulae, which is supported by observations.

   \end{abstract}

%
%  14.Sep.'90: Demo-Vs.
%________________________________________________________________

\section{Introduction}

The formation of a Planetary Nebula (PN) is critically determined by
the star which lives and evolves in the middle of it. The star
produces both the wind and the photons which determine the shape and
appearance of the nebula. This is why the most advanced models for the
formation of PNe are combinations of hydrodynamic and photo-ionization
calculations and include the evolutionary track of the star (Marten \&
Sch{\"o}nberner 1991; Mellema 1994, henceforth M94; Mellema 1995,
henceforth M95; Steffen et al.~1997).

Calculations of the evolution of stars after the Asymptotic Giant
Branch (AGB) phase show that this strongly depends on their mass (see
{e.g.\ }Vassiliadis \& Wood 1994; Bl{\"o}cker 1995), and one might expect
that this will have an effect on the formation of the nebula. A
typical post-AGB track in the HR-diagram consists of two parts
(Paczy\'nski 1971). Initially the star contracts, evolving to higher
effective temperatures at a constant luminosity; then as the energy
production stops, the luminosity starts decreasing and the effective
temperature starts dropping. The star reaches its highest temperature
(several times $10^5$~K) at the transition from the contraction to the
cooling phase.

In the contraction phase the more massive stars are more luminous and
can reach higher effective temperatures than the lower mass ones. The
most striking difference is however the evolutionary time scale. The
lowest mass stars can take several 10\,000~years to reach their
highest temperature, whereas the highest mass ones do the same in less
than a 100 years. This has several interesting effects, such as a low
probability to find more massive central stars in the contraction
phase.

Because the stellar radiation determines to a large extent the
appearance of the nebula, one would also expect the shape of the
nebula to depend on the evolutionary track of the central star. Here I
report the first, preliminary, calculations of this effect. To study
this I use a radiation-hydrodynamics code which follows the formation
of cylindrically symmetric PNe. This code has been tested and used
extensively to study the formation of aspherical PNe.

In Sect.~2 I give a short description of the numerical method and
initial conditions. Section~3 contains a description of the results,
which are further discussed in Sect.~4. The conclusions are summed up
in Sect.~5.

%__________________________________________________________________

\section{Model and code}

The formation of the PN is modelled using the Interacting Stellar
Winds (ISW) model. It is assumed that during the AGB phase the star
lost a large amount of material with a cylindrically symmetric density
distribution. As the star heats up during the post-AGB phase a fast
wind starts sweeping up this AGB material, shaping a PN out of it. The
AGB material is supposed to be denser in the equatorial plane (because
of the presence of a binary companion or perhaps some other effect,
see {e.g.\ }Livio 1997).  Models of this type have been successfully
used to explain aspherical PNe (see {e.g.\ }Frank \& Mellema 1994;
Mellema \& Frank 1995ab; M95).

The code used to produce the models is a combination of a 2D
hydrodynamics code (an approximate Riemann solver, described in
Mellema et al.~1989) and a photo-ionization/cooling calculation. Each
time step the heating due to photo-ionization and cooling due to the
most important permitted and forbidden lines, as well as free-free
radiation, are calculated. At the same time the time-dependent
ionization fractions are calculated. This method is described in more
detail in Frank \& Mellema (1994a).

The initial conditions are the same as described in Frank \& Mellema
(1994b), Mellema \& Frank (1995) and M95. In these the asphericity of
the AGB wind is characterised by two parameters: the density contrast
(the ratio between the density at the equator and at the pole), called
$q$ and a shape parameter $B$ (in some of the previous work called
$\beta$). For the two simulations presented here I used the same
initial conditions for the AGB wind, see Table 1. The parameter $B$
has the low value of 0.5, meaning that the AGB material is highly
concentrated towards the equator. In the pure ISW model without
stellar evolution effects this leads to the formation of a bipolar
PNe, see {e.g.\ }Frank \& Mellema (1994b).

The central star's evolution is taken from the work of Bl{\"o}cker
(1995). I use the tracks for 0.605 and 0.836{\msun} remnants. From the
star's luminosity and effective temperature I calculate the fast wind,
using the prescription from Pauldrach et al.~(1988) based on the
radiation pressure on lines. This procedure is the same as in M94 and
M95. The exact parameters for the star and the AGB mass loss are
listed in Table~1.

%__________________________________________________ One column table
   \begin{table}
      \caption{Parameters for runs A and B}
         \label{KapSou}
      \[
         \begin{tabular}{lll}
            \hline
            \noalign{\smallskip}
            Run & A & B \\
            \noalign{\smallskip}
            \hline
            \noalign{\smallskip}
            $\mdot_{\rm AGB}$ (\myr) & $1\,10^{-5} $ & $1\,10^{-5} $ \\ 
            $q$ & 5 & 5 \\
            $B$ & 0.5 & 0.5 \\
            $v_0$ (m s$^{-1}$) & $1.5\,10^{4} $ & $1.5\,10^{4} $ \\
            $T_0$ (K) & $2.0\,10^{2} $ & $2.0\,10^{2} $ \\
            stellar track (\msun) & 0.605 & 0.836 \\
            $r_0$ (m) & $1\,10^{14} $ & $1\,10^{14} $ \\
            $\Delta r$ (m) & $1.5\,10^{13} $ & $1.5\,10^{13} $ \\
            grid dimension & $100\times100$ & $100\times100$ \\
            \noalign{\smallskip}
            \hline
         \end{tabular}
      \]
   \end{table}

\section{Results}

%                                     
%______________________________________________ 
\begin{figure*}
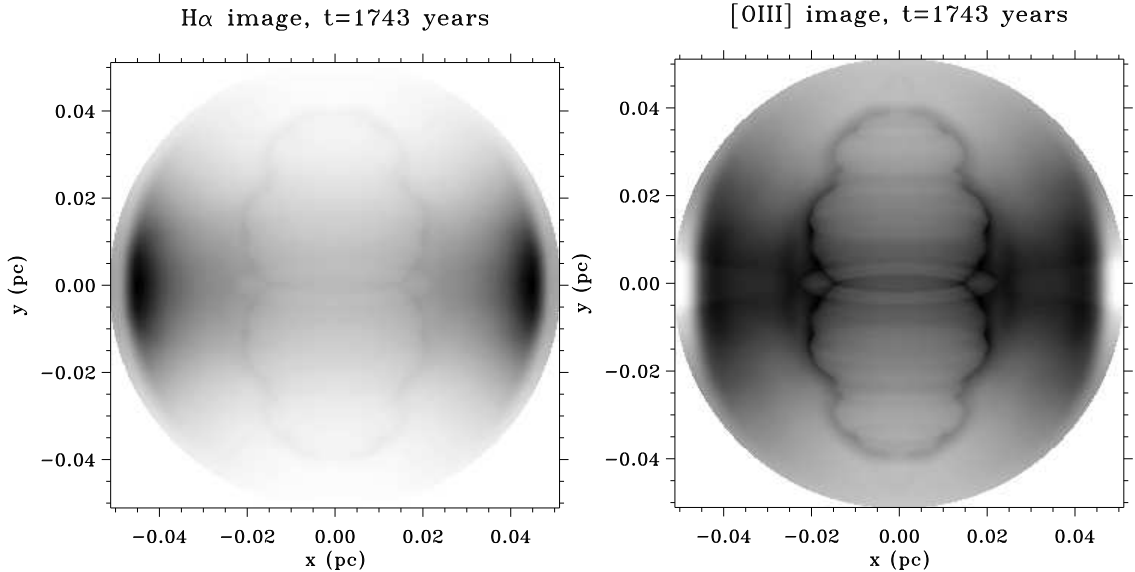

\centerline{\psfig{figure=imrh605_10.epsi,width=7.5cm}%
\psfig{figure=imo2605_10.epsi,width=7.5cm}}
\caption{Images in H$\alpha$ (left) and [O~III] (right) for run A}
\label{FigGam}%
\end{figure*}

Here I present the results of the two simulations. Figures 1 and 2
show some synthesized narrow-band images from these two
simulations. These images were constructed by taking the
two-dimens\-ional output of the code and rotating it around the
symmetry axis before projecting it on the sky. The angle at which all
images are shown is 10$^\circ$.

As was described in M95, the evolution in these types of models is
quite complex because of the time dependence of both the stellar
spectrum and the fast wind. The snapshots in the figures do not really
do justice to the amount of information produced by the simulations.
Nevertheless some trends can be discerned.

Run A displays the double shell evolution described in M95. Initially
a shell is swept up by the H ionization front. It is this structure
(called the I-shell in M95) which shows up most prominently in the
H$\alpha$ image. At the same time the fast wind is shaping a shell
(called the W-shell in M95) inside of this, and it is this shell which
shows up in the [O~III] image. Figure~1 shows the situation at
1743~years after the end of the AGB when the star has reached an
effective temperature of $28\;000$~K. The shape of both the I- and
W-shell is more or less elliptical. As time goes by the W-shell will
become the brighter one in all lines, but here I only followed the
evolution until the time the nebula reached the edge of the
computational grid.

Run B displays a very different behaviour. Here the ionization front
does {\it not} produce a shell and the wind shapes a bipolar
(butterfly) type of shell. This shell is most clearly seen in the
H$\alpha$ image. Because of the very high effective temperature of the
star the [O~III] image shows more of the surrounding material. But the
morphology of both images is much more bipolar than the ones from run
A. At a time of 507~years after the end of the AGB, the stellar
effective temperature is $226\;000$~K.

\section{Discussion}

Both models are still in their early evolutionary stages and one
should therefore be careful drawing far reaching conclusions, but it
is clear that model A is on its way to become an elliptical double
shell PN, whereas model B is tending to a more bipolar shape.

The physical reason behind this is the different time scales on which
the central stars evolve. For the most massive star the increase in
the number of UV photons is so fast that the ionization front never
stalls to become a D-type front. In other words: the ionization is
almost instant. This means that the original equatorially condensed
density structure is not altered appreciably by the ionization
process. In the 0.605{\msun} case, however, the ionization front
progresses slowly as a D-type front, having a shock front running
ahead of it. It is this D-front which produces the double shell
structure. This is likely to be the explanation for the `attached
haloes' (Pasquali \& Stanghellini 1995), as was already shown in M94.

M95 showed that the D-front can modify both the radial and tangential
density distributions as it moves out. In the case shown in M95, the
asphericity profile was smeared out somewhat, but this did not lead to
a clearly different morphology because the initial density
distribution was not very confined ($B=1$). For the lower value of 0.5
used here, the change after the passage of the front is
non-negligible. In run A the `effective $B$' after the passage of the
ionization front has been raised to about 1, leading to the formation
of an elliptical nebula, whereas one would have expected a bipolar
shape if there had been no modification by the ionization front. The
surprise here is not that the high mass star forms a bipolar nebula,
but rather that the low mass star does not!

This effect does not rule out the formation of bipolar PNe around
lower mass stars, but it does make it harder for them to form and thus
introduces a bias of bipolar shapes towards the more massive
precursors. There is good observational evidence that bipolarity
correlates quite well with stellar mass. Corradi \& Schwarz (1995)
studied a large sample of PNe and compared the properties of bipolars
with those of elliptical PNe. Their conclusion was that the bipolar
PNe form a separate group. They have on average a lower scale height
in the galaxy, higher N abundance, and their central stars have higher
effective temperatures. From this they conclude that on average the
central stars must have higher masses. These conclusions were
confirmed by G{\'o}rny et al.~(1997). Soker \& Livio (1994) suggested an
explanation in the frame work of a common envelope binary, and the
results in this paper provide another one which can also be used to
explain why not all known close binaries have bipolar PNe around them
(Bond \& Livio 1990).  Stanghellini et al.~(1993) did report a flat
distribution of central masses for bipolar nebulae, but this was based
on a smaller sample than the ones listed above.

The evolution described above is of course not the only possible
one. In general one would expect a more massive star to have had a
higher mass loss rate during the AGB, and thus a higher slow wind
density. This effect is enforced by the fact that during the
contraction phase the AGB material will be closer to the star, and
hence less diluted. This material will thus be less easily
ionized. But a more massive star is also more luminous and produces
more ionizing photons, so these two effects may cancel. The
observational evidence suggests that the densest parts of the AGB wind
do remain largely neutral since bipolar nebulae often show molecular
tori and effects of dust obscuration near the `waist'.

However, whether the AGB wind around these massive stars gets ionized
or not is not essential for the formation of the bipolar nebula. The
reason for this is that the initial evolution is fast. If an
ionization front forms it will quickly reach its largest size and then
as the star cools down, retreat again, or at best follow the
wind-swept shell. The density distribution in the AGB wind will not be
affected much by this and it is this which determines the final shape
of the nebula.

The results in this paper do not mean that more massive stars will
necessarily have bipolar nebulae around them. Whether or not this
happens still depends on the mass loss geometry of the AGB wind and
hence on the mechanism which produces the aspherical mass loss. If the
mass loss is not strongly concentrated towards the equator, a bipolar
PN will not form.

Note that the fact that symbiotic systems often also have bipolar
nebulae around them does not support or contradict the explanation
outlined above since it is difficult to say whether ionization is
rapid or gradual in these systems.
%                                     
%______________________________________________ 
\begin{figure*}
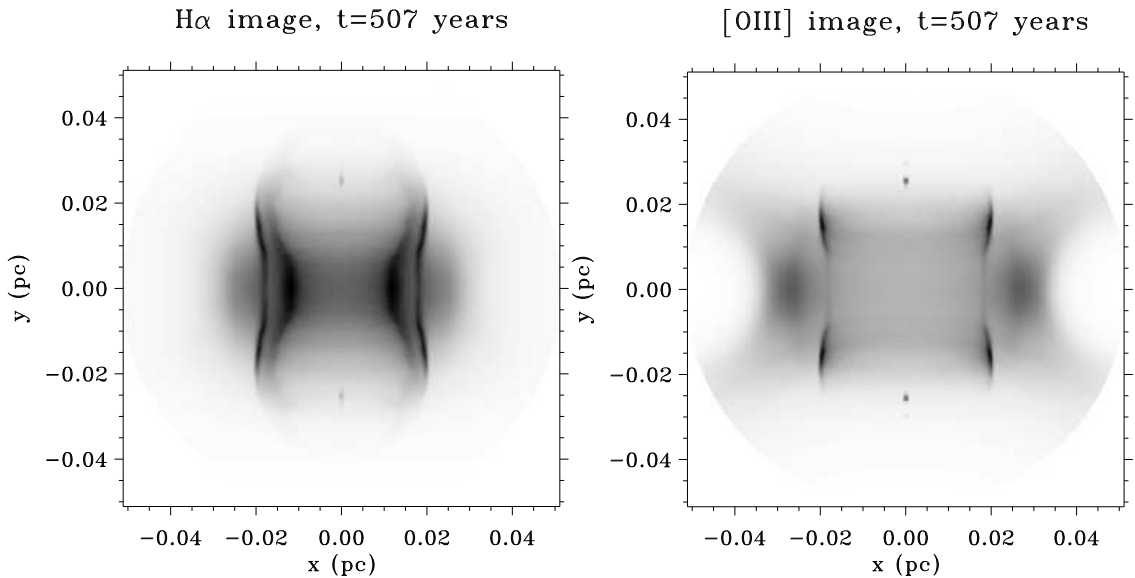

\centerline{\psfig{figure=imrh836_10.epsi,width=7.5cm}%
\psfig{figure=imo2836_10.epsi,width=7.5cm}}
\caption{Images in H$\alpha$ (left) and [O~III] (right) for run B}
\end{figure*}

\section{Relation between AGB mass loss and PN shape}

One conclusion from the ISW model is that the AGB wind density
distribution determines the shape of the PN. This principle was used
recently by Soker (1997) to derive a classification of PNe according
to the processes which caused their progenitors to have an
axi-symmetric AGB mass loss. Although there are a lot of assumptions
that go into a classification like this, it is still useful to have
it, if alone to focus further research.  But the results in this paper
show that one should be careful with trying to derive mass loss
properties from PN morphologies. It shows that the same AGB mass loss
geometry can result in very different morphologies for the
PN. Depending on the mass of the central star the PN acquires either a
bipolar shape or an elliptical with attached halo morphology. Other
numerical studies of PN formation around evolving low mass stars also
show the formation of attached haloes (Marten et al.~1994; M94; M95;
Steffen et al.~1997). So, it appears that especially the presence of
these attached haloes is an indication of a modification of the
original density distribution. The observational result that almost
all PNe with attached haloes are elliptical is fully consistent with
this (Stanghellini \& Pasquali 1995).

\section{Conclusions}

\begin{enumerate}

\item Numerical hydrodynamic models support the connection between
massive stars and PNe with a bipolar morphology. The models show that
it is more difficult for lower mass stars to acquire a PN with bipolar
morphology, even if the initial conditions are favourable.

\item The reason for this is that for lower mass stars the ionization
front is more likely to modify the surrounding density distribution,
making it less equatorially concentrated. To produce bipolar nebulae
one needs a density distribution with a high degree of concentration
towards the equator.

\item Around more massive stars the original density structure is more
likely to remain intact because either the material is so dense that
it remains neutral and the ionization front lies within the wind swept
shell, or the ionization is almost instantaneous (because of the rapid
evolution) and does not modify the density structure.

\item Since an equatorially concentrated, disk-like density
distribution is required for the formation of bipolar shapes, any
mechanism giving rise to aspherical mass loss on the AGB should be
able to produce this type of density distribution.
\end{enumerate}

%\begin{acknowledgements}
%\end{acknowledgements}

\end{document}